# Near-membrane refractometry using supercritical angle fluorescence

Maia Brunstein,*,† Lopamudra Roy,*,†,‡ ¶,§ and Martin Oheim*,†

*CNRS UMR 8118, Brain Physiology Laboratory, Paris F-75006, France; †Fédération de Recherche en Neurosciences FR 3636, Faculté de Sciences Fondamentales et Biomédicales, Université Paris Descartes, PRES Sorbonne Paris Cité, F-75006 Paris; ‡Erasmus Mundus International Master Europhotonics-POESII; ¶Karlsruhe School of Optics and Photonics (KSOP), Karlsruhe Institute of Technology (KIT), Karlsruhe D-76131, Germany; §Aix Marseille University, CNRS UMR 7249, Centrale Marseille, Institut Fresnel, Marseille F-13013, France.

ABSTRACT   Total internal reflection fluorescence (TIRF) microscopy and its variants are key technologies for visualizing the dynamics of single molecules or organelles in live cells. Yet, truely quantitative TIRF remains problematic. One unknown hampering the interpretation of evanescent-wave excited fluorescence intensities is the undetermined cell refractive index (RI). Here, we use a combination of TIRF excitation and supercritical angle fluorescence emission detection to directly measure the average RI in the 'footprint' region of the cell, during imaging. Our RI measurement is based on the determination on a back-focal plane image of the critical angle separating supercritical and undercritial fluorescence emission components. We validate our method by imaging mouse embryonic fibroblasts. By targeting various dyes and fluorescent-protein chimerae to vesicles, the plasma membrane as well as mitochondria and the ER, we demonstrate local RI measurements with subcellular resolution on a standard TIRF microscope with a removable Bertrand lens as the only modification. Our technique has important applications for imaging axial vesicle dynamics, mitochondrial energy state or detecting cancer cells.

Total internal fluorescence (TIRF) has evolved from an ex-pert technique to a routine contrast mode used for single-molecule and single-organelle tracking at or near the basal plasma membrane of cells adherent to a glass substrate [1]. PALM/STORM localization-based super-resolution micro-scopies have further broadened the range of TIRF applica-tions. The emission counterpart of TIRF, supercritical angle fluorescence (SAF) [2] is increasingly being used for sur-face microscopies [3, 4, 5]. For TIRF, the presence of a re-fractive-index (RI) boundary between the glass substrate (of index $n_2$) and the aqueous sample ($n_1$) results in the emergence of an evanescent wave that provides excitation-light confinement. In SAF microscopies, the otherwise un-detected evanescent emission component of surface-proxi-mal fluorophores couples to the interface where it becomes propagative and detectable in the far field, provided the ob-jective has a sufficiently large numerical aperture (NA= $n_2 \sin\theta_{NA}$, with $\theta_{NA} > \theta_c$). The critical angles at the excitation or emission wavelength $\lambda$, $\theta_c = \mathrm{asin}[n_2(\lambda)/n_1(\lambda)]$, the axial decay length $\delta = \lambda/[4\pi(n_2^2 \sin\theta - n_1^2)^{1/2}]$ of the evanescent-wave intensity and the dipole radiation pattern are all modi-fied by the *local* sample RI $n_1(x,y)$, which is generally un-determined. Knowledge, even of the *average* near-mem-brane RI, $\langle n_1 \rangle$, for the very cell under study, would greatly enhance our capacity to chose appropriate incidence and detection angles, to better understand and eliminate image imperfections and to interpret TIRF and SAF in quantita-tive terms, e.g., for axial profilometry, size- or concentra-tion measurements, or for axial single-vesicle tracking.

Hilbert-phase microscopy [6], digital confocal microscopy (DCM) [7, 8] and full-field optical coherence tomography (OCT) [9] all allow RI measurements with subcellular reso-lution but none of these techniques selectively probes the near-membrane space. On the other hand, DCM in a TIR geometry [10] and surface-plasmon-based RI sensing [11] both can, in principle, measure $\langle n_1 \rangle$ near the basal plasma membrane, but they require either a reference beam (and hence modifications to the TIRF illuminator) or else only work with metal-coated substrates or NSOM probes [12].

In this letter, we present a simple scheme for near mem-brane refractometry only requiring a Bertrand lens (BL, **Fig. 1**A). We measured $\langle n_1 \rangle$ of different organelles in the region probed by TIRF and SAF microscopies in live cells.

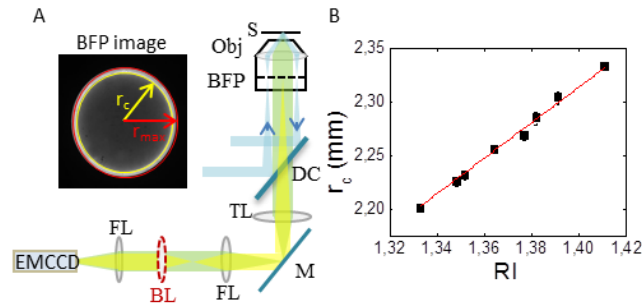

FIGURE 1. *SAF-based sample-refractive index measurement*. (A) Simpli-fied optical layout. Obj – objective, *BFP* – backfocal plane, TL – tube lens, BL – Bertrand lens, FL – focusing lens, EMCCD – electron-multi-plying charge-coupled device, S- sample, DC- dichroic mirror, M- mirror. Excitation wavelength was 488nm. *Inset* shows BFP image of a thin layer of 488/515nm (100-nm diameter) beads in water. $r_c$ identifies critical angle at a given RI. $r_{max}$ corresponds to the objective maximum angle of collection. (B) RIs measured by Abbe refractometry from a series of cali-brated sucrose standards vs. the corresponding $r_c$. Red line is linear fit of the data, $f = (1.6528 \pm 0.0003)n_1$. $R = 1$. See **SI online** for details.

Our technique is based on the realization that for fluoro-phores located near the reflecting interface and embedded in a medium of RI ($n_1$) this value is encoded in the aperture plane (backfocal plane, BFP) of the objective as the boun-dary separating undercritical and supercritical fluorescence emission components (Fig. S1). Their limit is the critical angle $\theta_{c,fluo}$ at the wavelength of fluorescence emission. In the BFP, angles are converted to radii, and the sample RI and objective NA correspond to $r_c = f \cdot \langle n_1 \rangle$ (*yellow* line on the *inset* in **Fig. 1***A*) and $r_{max} = f \cdot n_2 \cdot \sin\theta_{NA} = f \cdot NA$ (*red*) re-spectively, where $f$ is the objective focal length. We found an effective $NA_{eff}$ of $1.465 \pm 0.009$, consistent with earlier measurements of the lens [13]. $f_{eff} = 1.6528 \pm 0.0003$ was obtained from a line-fit of the known RIs of a series of calibrated sucrose standards with the measured radii on the corresponding BFP images, in excellent agreement with the nominal focal length for the ZEISS $\alpha$PlanApo x100/1.46 NA-objective (nominal $f=1.65$), **Fig. 1***B*.

Our technique probes the local RI of the medium in which SAF-emitting fluorophores are embedded. We there-fore reasoned that RI measurements with subcellular reso-lution should be possible by targeting the fluorophores to specific subcellular compartments (**Fig. S2**). We measured, using AOD-based spinning TIRF [14] excitation (at $\theta = 68°$) the RIs of different organelles in mouse embryonic fibro-blasts (MEFs) with fluorophores. **Fig. 2** compiles the RIs obtained from BFP images of near-membrane endo-/lyso-somes (labeled with FITC-dextran, FM2-10 or FM1-43, respectively), mitochondria (labeled with MitoTracker-Green or Mito-EGFP) and the endoplasmic reticulum (ER, labeled with ER-EGFP). The *inset* shows the BFP image of a MEF labeled with the amphiphilic dye FM1-43. We found cellular $\langle n_1 \rangle$ values ranging from 1.344 to 1.4 ($n_{water} = 1.335$ at 510 nm) [14]). These cellular RI measurements were robust for different regions of interest, and across cells, lending further support to the approach (**Fig. S3**).

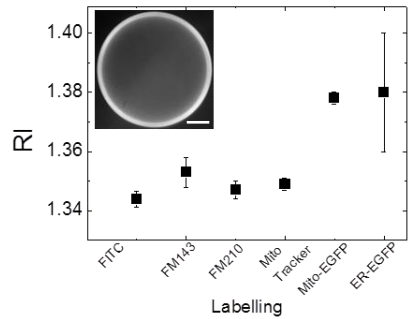

FIGURE 2. *Subcellular refractive-index measurements*. SAF-based RI measurements in live MEFs for different subcellular stainings. Error bars correspond to the SD of the mean RI for 6-8 different cells each. *Inset* shows the BFP image of a MEF labeled with FM1-43 (a vesicular stainging, predominantly of endo-/lysomes in MEFs). Scale bar: 1 mm.

Does the dense protein packageing at cell adhesion sites af-fect the RI of the near-membrane layer compared to dee-per cytoplasmic layers? To adress this question, we mea-sured in MEFs $\langle n_1 \rangle$ for different polar angles, $\theta = 68°$ and $73°$, and therefore different EW penetration depths $\delta$. We observed no systematic trend (**Fig. S4**). The lack of a clear $z$-dependence of the RI can be understood when consi-dering the detailed ring structure on BFP images, **Fig. 3**. For FM dyes, we observed a well-contrasted bright SAF ring, as the fluorophores were located in the basal plasma membrane and near-membrane vesicles, close to the reflec-ting interface, **Fig. 3**, *left*. In contrast, for a mitochondrial or ER-labeling, fluorophores are located more distant from the interface and their emission is dominated by UAF (*right*), see [15] and **SI Discussion**. As a corollary, we can conclude from this observation, too, that the excitation light was not perfectly evanescent but also presented long-range (scattered light) components as already discussed in our earlier work using the same microscope [13].

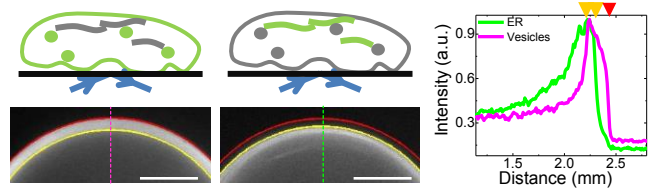

FIGURE 3. *Influence of axial fluorophore localization on the BFP image*. (*Top*) Sketches of vesicular and plasma membrane labeling (*left*), as well as reticular labeling in MEFs (*center*). Note the difference in fluorophore distance $z$ from the interface. (*Bottom*) Details of the corresponding BFP images showing the fitted $r_{max}$ (red line) and $r_c$ (yellow line). Scale bar is 1 mm. (*Right*) Graph on the *right* shows intensity profiles along the purple and green lines on the BFP images, and the fitted $r_{max}$ and $r_c$ values with red and yellow arrows, respectively.

Evidently, the quality of our SAF-based refractive-index measurements depends on the accuracy and precision of measuring radii in the objective BFP, which in turn de-pends on the pixel calibration of the BFP image and the way radii are extracted. We re-normalized the BFP image pixel size with the known RI values of air and water, re-spectively (see **SI Materials**). Radii were fitted as descri-bed [13], (**Fig. S4**), and the excellent agreement between the apparent and true objective focal length (**Fig. 1**) war-rants the validty of our method. Overall, we can estimate a 0.2% uncertainty in precision of RI.

A fundamental limit of our technique is that it reports the *average* RI of the local environments embedding all (near-surface) fluorophores across the entire field-of-view. The presence of membrane-distant fluorophores that cannot emit SAF renders the detection of the $r_c$ ring more difficult. As such, fluorophore targeting and/or excitation-light con-finement improve the spatial resolution of the measure-ment, as would, e.g., photoswitching and imaging of fluorophore subpopulations or confocal spot detection [15].

Finally, albeit trivial, fluorophore brightness and concen-tration must exceed instrument and sample autofluore-scence that eventually limit the sensitivity of our method (Bernhard, Brunstein & Oheim, *in preparation*).

SAF-based refractometry is not only compatible with the now preferred azimuthal angle scanning TIRF [14, 16-19] as used here, but also with confocal spot scanning TIRF [15], HILO [14, 20], and, at least for thin samples, epifluo-rescence. Combining inexpensive LED illumination SAF-refractometry in a disposable biosensor our technique will allow, e.g., low-cost on-chip diagnosis of cellular mali-gnancy [21] or mitochondrial bioenergetics [22].

**ACKNOWLEDGEMENTS**

We thank Gérard Louis (Paris) for the loan of the Abbe refractometer, the Hirrlinger (Leipzig) and Estaquier labs (Paris) for help with molecular biology, and Dan Axelrod (Ann Arbor), Rodolphe Jaffiol (Troyes) and Gilles Tessier (Paris) for discussions. Patrice Jegouzo (Paris) and Wolfram Lessner (Göttingen) provided custom mechanics. We thank Maria Camila Tovar Fernandez, Isabelle Nondier and Camilla De Fazio for MEF cell culture. Support by the large-scale National Infrastruc-ture France-BioImaging (FBI, ANR-10-INSB-04, Invest-ments for the future), the CNRS (*Défi instrumentation aux limites*) and the European Union (H2020 JPND Synspread). The Oheim lab is a member of the Federation of Neuroscience labs (CNRS FR3636), of the C'nano IdF (CNRS GDR2972) and the Ecole de Neurosciences de Paris (ENP) excellence clusters for nanobiotechnology and neuroscience, respectively.

**AUTHOR CONTRIBUTIONS**

MB, LR and MO performed experiments and analyzed data, MO conceptualized and directed research and wrote the manuscript with contributions from all authors.

**SUPPORTING MATERIAL**

## Supplementary figures

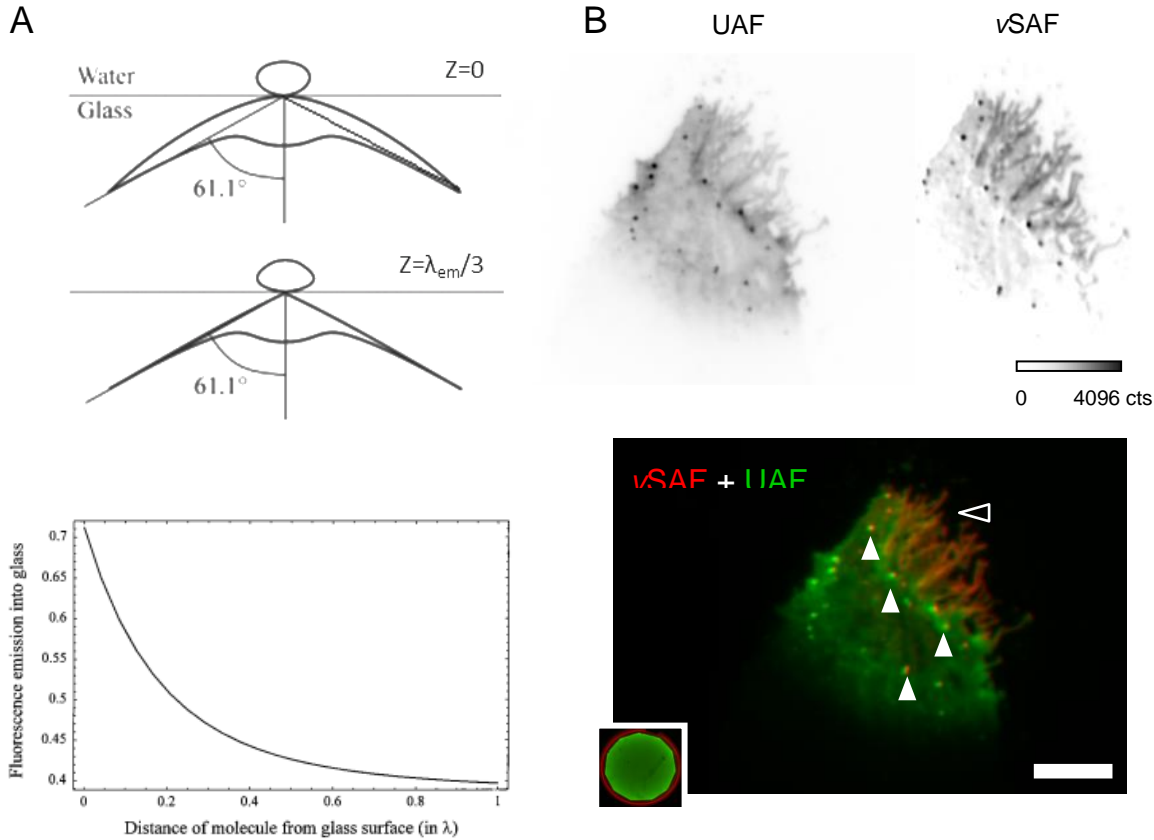

**FIG. S1** *Supercritical angle fluorescence (SAF) selectively probes the near-interface layer of cells cultured on a glass substrate.* (A), *top*, calculated radiation patterns (from ref. [1] for fluorophores located directly at the reflecting interface ($z = 0$ nm, *top*) and at 1/3 of the emis-sion wavelength $\lambda_{em}$ (*bottom*). Dipole moments are assumed to be oriented isotropically. *Bottom*, Integrated fractional intensity of dipole radiation emitted into the lower half space as a function of the axial fluorophore distance $z$ (from ref. [2]). Note the steep (near-) exponent-tial intensity loss, due to the surface coupling of evanescent dipole emission components that become propagative and are collected by the high-NA objective. (B), emission optical sec-tioning by angular filtering. Undercritical-angle fluore-scence (UAF, *top*) and virtual SAF (*v*SAF) images (see [3, 4] for experimental details) of a mouse cortical astrocyte expressing CD63-EGFP, and their red/green pseudo-color overlay (*bottom*). Note the adherent leading edge of the cell (*open arrowhead*), close to the glass substrate, as well as near-membrane endo-/lysosomes that moved between SAF and UAF acquisitions (*solid arrowheads*), $\lambda_{ex}/\lambda_{em} = 488/515$ nm. Exposure time was 50 ms. Scale bar, 5 μm. *Inset* shows the pseudocolor over-lay of the SAF (*red*) and UAF (*green*) BFP images. The UAF Fourier-plane image also shows the facets of the iris blocking the SAF ring during UAF acquisitions. Objective was a ×100 /1.46 α-PlanApo-chromat (ZEISS). See **Supplementary Discussion**.

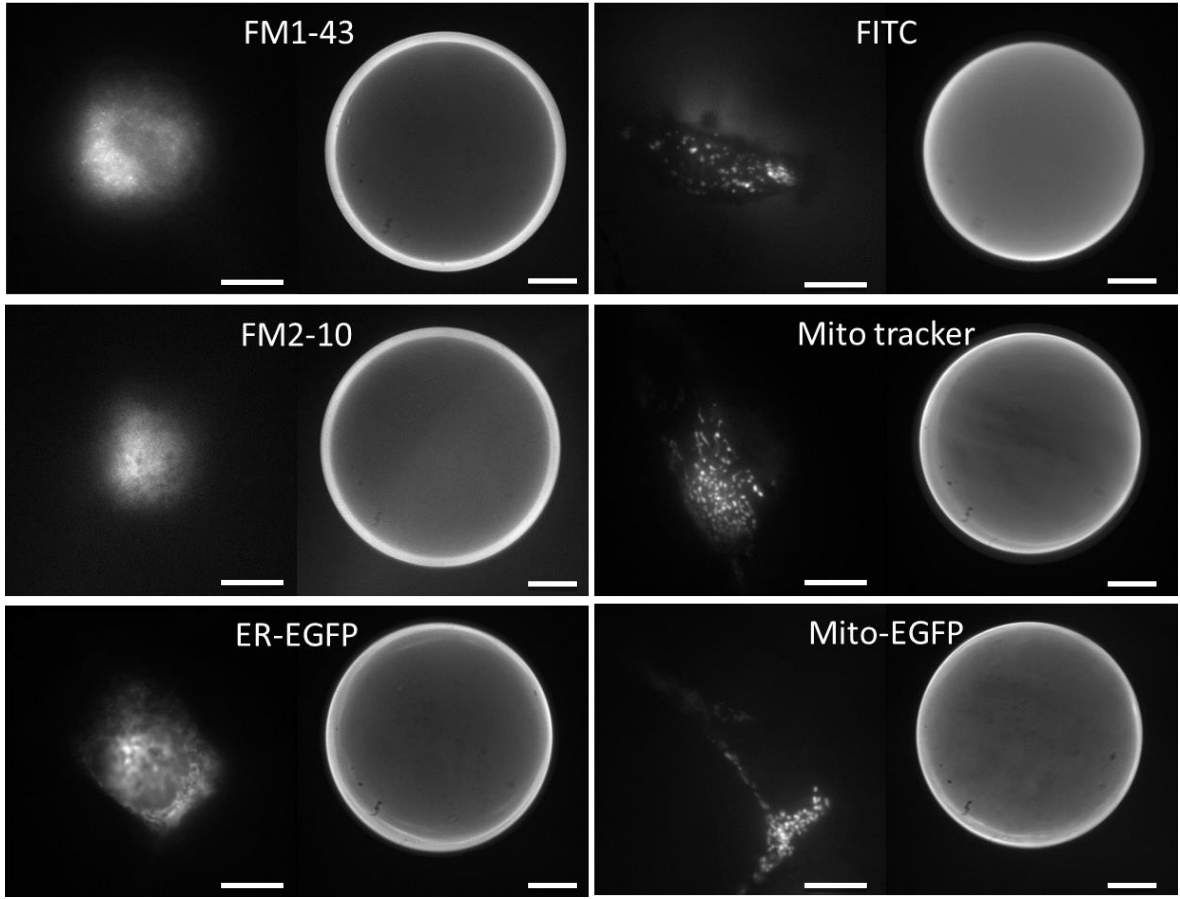

**Fig. S2.** *Specificity of subcellular stainings.* We systematically acquired sample- and Fourier-plane images by alternating images without and with Bertrand lens, respectively. *Left* on each panel, typical sample-plane spTIRF images (upon excitation at $\lambda_{ex}$ = 488 nm and $\theta$ = 68°) of mouse embryonic fibroblasts (MEFs) with either vesicular and plasma membrane labeling (FM1-43, FM2-10), vesicular labeling (FITC dextran), or mitochondrial labeling following expression of Mito-EGFP, or incubation with MitoTrackerGreen or, finally, ER labeling after expression of a ER-targeted fusion construct (ER-GFP). *Right* images on each panel show the corresponding raw-data aperture-plane (BFP) images that were used for the RI measurement. Scale bars, 10 µm (1 mm) for sample- (aperture-) plane images, respectively.

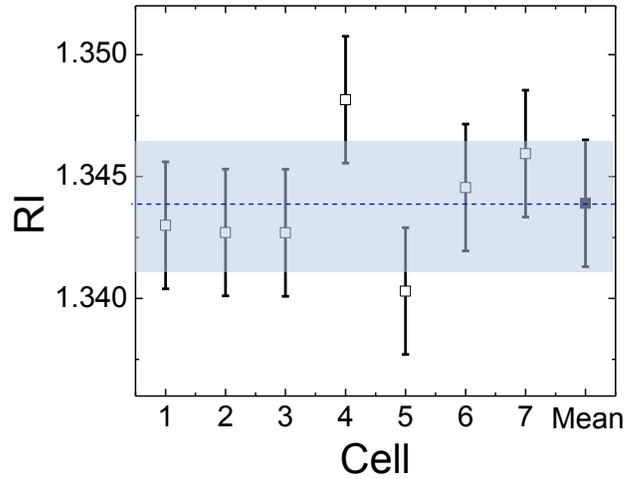

**Fig. S3**. *Robustness of SAF-based single-cell refractometry*. SAF-based refractive index (RI) measurement for *n* = 7 different MEFs labeled by spontaneous internalization of FITC-dextran (10,000 MW, 1mg/ml, 2 h, producing a vesicular endo-/lysosomal labeling through pino-/endocytic mechanisms). Note that the RI values found are in the same range. Open symbols and error bars are individual measurements and the corresponding precision of our method (0.2%, limited by the positioning of the Bertrand lens (BL), from three independent experiments with the BL repositioned each time) as well the uncertainty of the fitted $r_c$ and $r_{max}$. Solid symbol and error bar are population average and SD, indicative of the accuracy of the RI measurement (again of the order of 0.2%). *Blue* line and shade indicate average and SD, respectively.

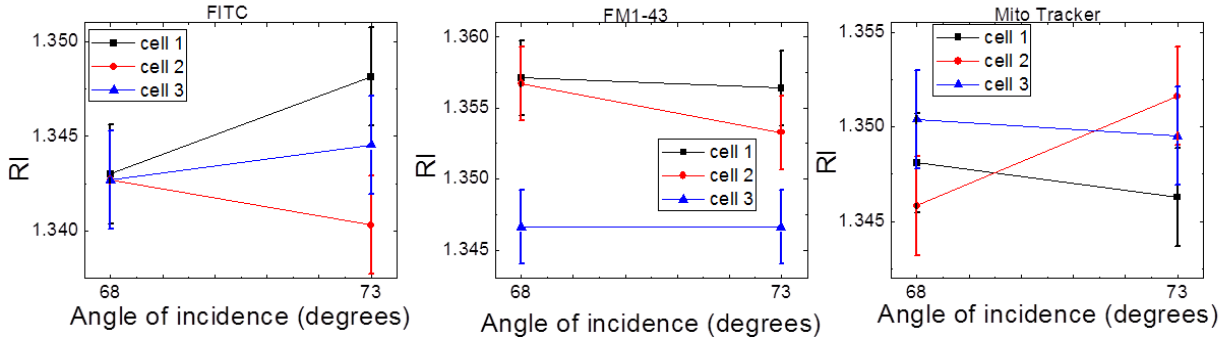

**Fig. S4**. *No systematic dependency of the RI on the polar angle of incidence and EW penetration depth*. SAF-based RI measurement from three different cells each with either vesicular (*left* and *middle*) or mitochondrial (*right*) staining, for two polar angles of incidence *θ* of the totally reflected 488-nm excitation beam, and hence two different penetration depths (of the order of ~90 and 70 nm, for 68° and 73°, respectively). No clear trend of the RI with penetration depth was observed. See **Supplementary Discussion** for details.

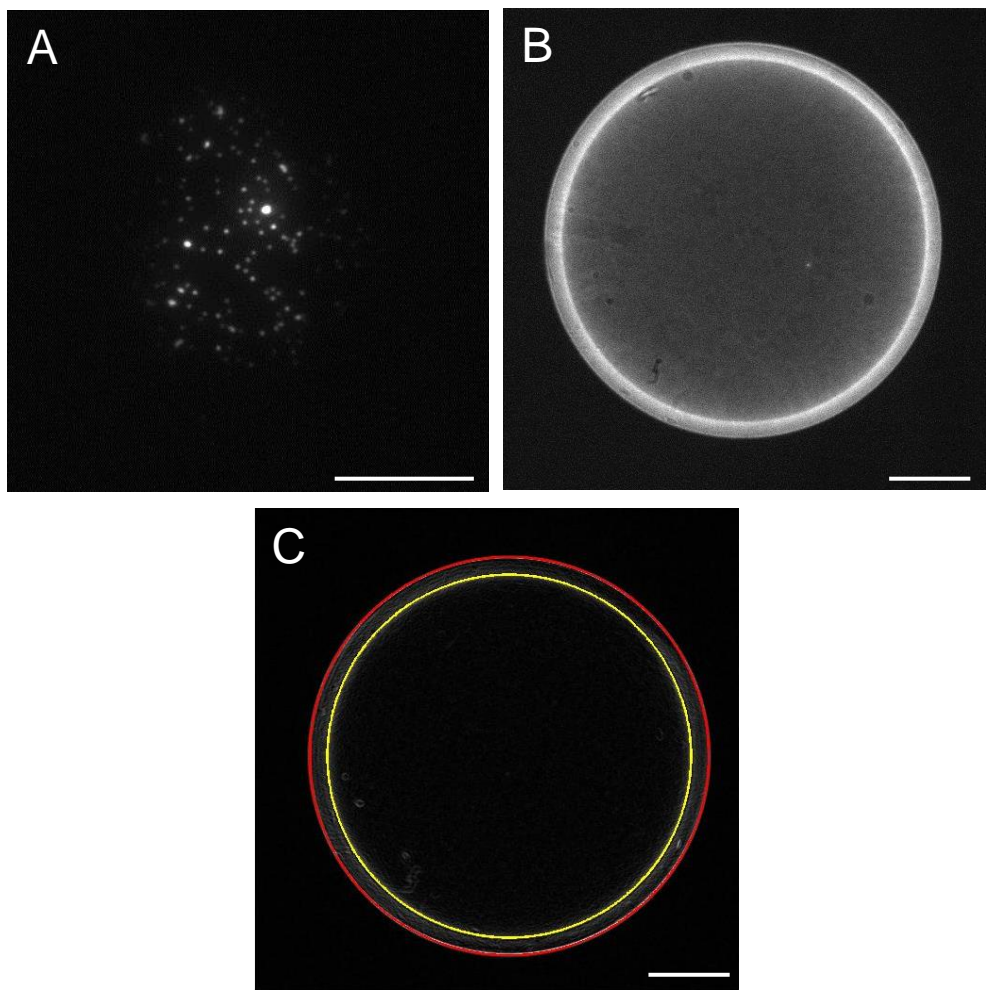

**Fig. S5**. *Fitting of $r_c$ and $r_{max}$*. (A), sample-plane image of a small drop of sub-resolution (100-nm diameter) fluorescent microspheres deposited on a borosilicate coverslip by solvent evaporation and, once dry, topped with water. Scale bar, 10 µm. (B), corresponding aperture-plane (Fourier-plane) image (radiation pattern), taken by the insertion of the Bertrand lens. Scale bar, 1 mm. Objective was the ×100/1.46 α-PlanApochromat (ZEISS) used throughout this study. (C), derivative of the BFP image in (B), from which $r_c$ (*yellow*) and $r_{max}$ (*red*) were determined by fitting circles with the intensity maxima (steepest slope) using a custom macro written in MATLAB. Scale bar is 1 mm as in panel (B).

## Supplementary tables

**Table S1**. *Used organelle markers*

|  | *endo-/lysosomes* | | | *mitochondria* | | *ER* |
|---|---|---|---|---|---|---|
| *Labeling* | FITC-dextran | FM2-10 | FM1-43 | Mito tracker green | Mito-EGFP (plasmid) | ER-EGFP (plasmid) |
| *Protocol* | 1mg/ml, 2h | 1µg/ml, 20' | 1µg/ml, 20' | 1µg/ml, 20' | 2µg/ml, 6h | 2µg/ml, 6h |

**Table S2**. *Measured refractive indices (RI) from calibrated sucrose standards*

| *Specified %* [a] | *Measured % ±0.25* [b] | *Offset (rel.)* | *Measured RI ±0.0005* [c] |
|---|---|---|---|
| 10 | 10.5 | 0,05 | 1.3485 |
| 12.5 | 12.75 | 0,02 | 1.352 |
| 20 | 20.5 | 0,025 | 1.3645 |
| 25 | 25.75 | 0,03 | 1.377 |
| 30 | 30.5 | 0,016 | 1.382 |
| 35 | 35.5 | 0,014 | 1.3915 |
| 45 | 45.5 | 0,01 | 1.411 |
| 50 | 50.5 | 0,01 | 1.4215 |
| 55 | 55.5 | 0,009 | 1.4325 |
| 60 | 61.75 | 0,029 | 1.4455 |
|  |  | *mean* : 0.02 |  |

[a] manufacturer-specified sucrose content with an uncertainty of measurement of 0.11% for all Brix values

[b] measured with an Abbe refractometer at 20°C and 589-nm excitation.

[c] SAF-based RI measurement from BFP images of 100-nm fluorescent microspheres topped with the respective sucrose standards. Excitation wavelength was 488 nm, $\theta = 68°$. Values are averages of three independent measurements.

**Table S3**. *Organelle refractive indices (RI) vs. literature values*

| *Mitochondria* | | *Endo-/lysosomes* | | *ER* | |
|---|---|---|---|---|---|
| 1.36 ± 0.02 | this study ($n$=10) | 1.348 ± 0.006 | this study ($n$=15) | 1.38 ± 0.02 | this study ($n$=8) |
| 1.4 | Ref. [5][a] | 1.6 | Ref. [5][a] | 1.549-1.601 | Ref. [23][c] |
| 1.41 ± 0.01 | Ref. [6][b] |  |  |  |  |

[a] Angularly resolved light scattering and wavelength-resolved dark-field scattering spectroscopy measurements on control EMT6 cells and cells stained with high-extinction lysosomal-or mitochondrial-localizing dyes.

[b] Isolated mitochondria in isotonic solution using retardation-modulated differential interference contrast microscopy.

[c] Measurements of light scattering of isolated subcellular fractionation ER in buffered glycerol-water mixtures.

## Supplementary experimental procedures

*Reagents.* DMEM cell culture medium (Gibco), FCS, penicillin/streptomycin and 100-nm diameter yellow-green emitting (505/515 nm) InSpeck[TM] beads were from Invitrogen (Cergy Pontoise, France). Fluorescein isothiocyanate (FITC) and FITC-dextran (FD-10S) were from Sigma (Deisenhofen, Germany). Beads were re-suspended in EtOH and deposited as a sparse monolayer on the coverslip surface by solvent evaporation of a small (µl) droplet. FM2-10 and FM1-43 were bath applied (1µg/ml, 20 min) and the cells thoroughly washed two times before imaging. FITC-dextran was applied at 1 mg/ml during 2h and the cells imaged after changing the extracellular solution, leaving only endo-/pinocytosed FITC in intracellular vesicles.

*Cell culture, plasmids and transfections.* For cell culture use, coverslips (BK-7, 25-mm diameter, Schott-Desag D263M, Thermo Fisher Menzel, Braunschweig, Germany) were sterilized by passing them through 70% ethanol (EtOH) baths, and kept individually in EtOH in a tightly sealed 6-well plate. Prior to use, the EtOH was removed, the coverslips rinsed with sterile water to avoid bacterial infections and covered with culture medium (see below). In some experiments, coverslips were pre-bleached by intense overnight 337-nm irradiation with 260-µJ pulses emitted from a pulsed nitrogen laser (VSL337ND-S, SpectraPhysics, Santa Paula, CA).

Mouse embryonic fibroblasts (MEFs) were cultured at 37°C in a 5% $CO_2$ atmosphere in a medium of 86% Dulbecco's modified Eagle's medium (DMEM), 9.6% of Fetal Bovine Serum (FBS), 1% of L-glutamine (200 mM, PAA Laboratories, Austria), penicillin/strepto-mycin (50 µg/ml), HEPES (pH 7.3, 1M), pyruvate (100 mM) and 0.01% *β*-mercapto-ethanol (50 mM, Gibco). For imaging, coverslips were shaken and rinsed to remove detached and faintly adherent dead cells, tightly fixed in a custom Teflon coverslip holder, and covered with imaging medium (the same as the culture medium but with DMEM containing no phenol red). All experiments were performed at room temperature (20°C).

*Refractive-index calibration.* RIs ($n_D^{(20°)}$) of pre-calibrated commercial sucrose 'Brix' stan-dards having 10, 12.5, 20, 25, 30, 35, 45, 50, 55 and 60 °Bx (i.e., %w/w sucrose in water, from Reagecon, Shannon, Ireland) were first verified at RT (20°C) with an Abbe refract-tometer (WYA, Shanghai, China) using the sodium-D line of a low-pressure sodium-vapor lamp (589 nm). They showed a systematic 2-% offset compared to the specified values, see **Table S2**. We used the same standards, as well as no medium (fluorescent beads in air) or water to top coverslips and collected the resulting fluorescence.

*Calibration of the pixel size on BFP images.* For pixel calibration, we placed a transparent microruler (Linos, Göttingen, Germany) in the in the BFP (8.5 mm above the objective shoulder) and measured the line spacing on BFP images. Pixel size was 11.5 µm. With this pixel size and using *f* =1.65 (the nominal value for a Zeiss 100x objective) with the followings equations $r_c = f \cdot \langle n_1 \rangle$ and $r_{max} = f \cdot NA$ we found values of RI for air ($n_{air}$= 1.003 ±0.003) and water ($n_{water}$= 1.339 ± 0.003) in excellent agreement with the true values. The objective NA found (NA=1.465±0.009) is also in good agreement with the nominal one (*NA* = 1.46).

*Combined TIRF-SAF microscopy.* The beam of an $Ar^+$-ion gas laser was filtered to remove the 458- and 514-nm lines, shuttered (LS3, Uniblitz, Vincent Assoc., Rochester, NY) and delivered to the optical table via a mono-mode optical fiber (Qioptiq PointSource, Hamble, UK). Polarization was adjusted with a zero-order half-wave plate (*λ*/2, WPH05M-488, Thorlabs, Newton, NJ) to maximize the throughput of the (1,1)-order of our AOD scanner (see below).

We used our combined spinning TIRF [7, 8] and virtual supercritical angle fluore-scence [4] (spTIR-*v*SAF) microscope described earlier. Briefly, a pair of acousto-optical deflectors (AODs, AA.Opto, St-Rémy-en-Chevreuse, France) scanned the expanded beam that was focused by six-element scan lens (Rodagon, Rodenstock, Feldkirchen, Germany) to a tight spot in the objective BFP. All other lenses were achromatic doublets (Linos). The $\alpha$Plan-Apochromat ×100/NA1.46oil oil objective (Zeiss, Oberkochen, Germany) was piezo-positioned for accurate focusing at the near-membrane layer (PIFOC, Physik Instrumente, Karlsruhe, Germany). Fluorescence was detected through the same objective that was used for TIRF excitation, extracted with a dichroic mirror (zt491 RDCXT, AHF, Tübingen, Germany), filtered with two stacked holographic notch filters (to remove the totally reflected 488-nm beam), and either directly imaged onto the camera or else the BFP imaged with a Bertrand lens ($f$ = 100 mm) onto an electron multiplying charge-coupled device camera (EMCCD, QuantEM512C, Photometrics, Tucson, AZ). The total magnification was 90.6 nm/pixel for sample-plane images and 11.5 µm/pixel for aperture-plane images.

Compared to the earlier published version [4], four modifications were made on our microscope: (*i*), a $\lambda$/4 plate (WPQ05M-488, Thorlabs) was inserted right after the AODs so as to produce a circular polarized 488-nm beam; (*ii*) the second lens of the scan-angle increasing (beam compressing telescope $f_1$ = 100 mm, $f_2$ = 50 mm) was mounted on a *z*-adjustment positioner (G061165000, Linos) and its exact axial position fixed by minimizing the beam divergence, measured as the spot diameter of an on-axis beam emerging from the microscope objective when projected at the ceiling; (*iii*), similar to ref. [7], we mounted a small inexpen-sive CMOS camera (DCC1545M, Thorlabs) in an equivalent back focal plane of the micro-scope excitation optical path to monitor the excitation light distribution; (iv), a white-light HighLED (Linos, ref. G06 5150 000, Luxeon Star LXHL-MW1C, 40-mW) provided bright-field illumination.

*Determination of $r_c$ and $r_{max}$.* In order to find the inner ($r_c$) and outer radius ($r_{max}$) form BFP images a custom routine was written in MATLAB (The Mathworks, Natick, MA) that first took the derivative along both *x*- and *y*-directions of the image and then fitted a circle with the maxima of the derivative image, see **Fig. S5**. As such, the accuracy of the fitting process is sub-pixel.

*Estimation of RI measurement accuracy and precision.* The accuracy of the method was measured by taking BFP images of a monolayer of sub-resolution fluorescent beads (100-nm diameter) covered with water or the glucose solution mentioned in Table S2 for three independent alignments (along *z*) of the Bertrand lens. The mean of the SD of the RI found for each solution was 0.0026, this error not only takes into account the alignment precision but also the accuracy of the fitting process. This value was considered as the error of all RI measurements.

**Supplementary discussion**

*Multilayer surfaces.* The cavity effects of multiple dielectric layers can produce complex resonances in the radiation pattern and multi-ring systems on the BFP image [9]. For cells cultured on a glass substrate, such layers result from polymer coating on the coverslip that is often used to favor cell adhesion, or they can be deposited by the cells themselves as a conse-quence of the secretion of cell-adhesion molecules or extracellular matrix proteins.

To test whether such multi-layer effects had to be taken into account for the interpret-tation of our BFP images, we took images of fluorescent microspheres on bare or poly-ornithine coated coverslips as well as on coated coverslips on which MEFs were grown during one week. We did not observe any significant differences in the emission pattern (not shown).

*Near-critical emission of surface-distant fluorophores.* Following the equations in refs. [10] and [11], we can calculate the angular emission pattern (radiation pattern) for a fluorophore located at an axial distance $z$ from the water/glass interface. Even at $z = 10\ \lambda_{em}$ ($\lambda_{em}$ being the emission wavelength) the radiation pattern (BFP image) still shows an intensity peak near the critical angle $\theta_{c,fluo}$ (not shown), however, this ring is observed for radii corresponding to subcritical polar angles, i.e., $\theta < \theta_{c,fluo}$. The difference with the radiation patterns of fluoro-phores located much closer to the surface is that *no* supercritical emission at $\theta > \theta_{c,fluo}$ is possible for distant emitters (this is because the surface-distant fluorophores cannot emit far-field light into these 'forbidden angles' and they are to distant from the surface for their near-field to couple to the interface and produce SAF) but they still have an intensity peak at very high UAF angles. As a consequence, even for distant fluorophores $r_{c,fluo}$ and hence $\theta_{c,fluo}$ and the RI can be obtained from the BFP image. The question is then if the ring intensity relative to the BFP image background due to UAF, instrument and cellular autofluorescence is still sufficient for precise edge detection.

*Influence of the excitation light distribution on the radiation pattern.* Depending on the pre-cise excitation-light confinement different fluorophore populations are excited: in TIRF, de-spite the presence of non-evanescent (long-range) components [8, 12], SAF-competent emit-ters dominate, but even these near-surface dipoles always emit a mix of SAF and UAF. On the contrary, for epifluorescence or confocal laser scanning microscopy both near-interface and distant fluorophores contribute to the BFP image, and because for the latter no near-field emission components can couple to the interface, distant fluorophores cannot emit light into forbidden angles. Surface-distant emitters contribute background (UAF emission) and will render the detection of the critical angle $\theta_{c,fluo}$ and hence the RI measurement more difficult. The ring intensity will be determined by the quantity, proximity and brightness ($\varepsilon\phi$) of near-interface fluorophores. Therefore, although SAF detection is – in principle – compatible with wide-field excitation, SAF-based refractometry will benefit from the additional excita-tion light confinement of TIRF. The analogous effect was already observed for SAF-based imaging of near-membrane organelles, see [8] and **Fig. S1**, where SAF was seen to further improve the axial optical sectioning of TIRF by adding an emission spatial filtering (Fourier-plane filtering). Thus, although both EW-phenomena - in theory - operate on the same length-scale (of the order of 100 nm), SAF can improve axial fluorophore localization and – in part - compensate for the imperfections of TIRF excitation.

In this context, our observation that the measured RI value does not systematically change when altering the penetration depth of the EW (**Fig. S4**) can be interpreted in three different manners. (*i*), the most straightforward interpretation is that there is no measurable difference in the local RI in the range probed by the two EWs. This would mean that on average, the local microenvironment of all SAF-emitting fluorophores would be the same in these zones, or at least that any difference – if there is any – was smaller than the error bar of the RI measurement; (*ii*), a second way of interpreting the data is that the subcellular locali-zation of the fluorophores to specific organelles out-weighs any *z*-effects. This would mean that their specific targeting exposes fluorophores to a local chemical microenvironment that merely reflects the organelle composition and that this environment is little affected by where precisely this organelle is located with respect to the basal plasma membrane. Indeed, despite some variation, the mean RIs over three measurements for the vesicular, membrane and mitochondrial RIs showed distinct and characteristic values for each type of organelle despite no significant difference at different beam angles argues; finally, (*iii*), yet another way of interpreting our data is that our technique relies on the detectability of the $r_c$ on BFP images and therefore, by definition, those fluorophores that are located closest to the coverslip surface will dominate the RI measurement. A 50-nm increase in the axial volume probed will change the radiation pattern, i.e., the intensity emitted as SAF and UAF, but it will not change $\theta_{c,\text{fluo}}$, (or $r_{c,\text{fluo}}$).

*Comparison with other subcellular refractive-index measurements.* While the importance of the cell RI for cell biology, biomedical imaging and disease diagnosis is well recognized (see, e.g., ref. [13] for a recent review), *local* RI measurements in live cells are fairly rare, and so are studies that specifically measure the RI in the near-membrane space. Due to their inherent surface selectivity, techniques that use optical effects at the dielectric boundary emerges as a natural choice: in their classical work, Bereiter-Hahn *et al.* [14] used quantitative interference reflection contrast (IRC) microscopy [15] to investigate how cell adhesion to the glass substrate locally modifies the RI. They found higher RI values (1.38 to 1.40) at points of focal contact, where stress fibers terminate, compared to areas of close contact (1.354-1.368). In areas of the cortical cytoplasm, between focal contacts, not adherent to the glass substrate, RIs between 1.353 and 1.368 were observed, however these numbers are likely to represent mixed values corresponding to the finite volume probed by IRC. Also, in the absence of a specific organelle labeling, IRC does not provide RI values for specific subcellular compartments. While the work of Bereiter-Hahn prompted our variable-angle TIRF (VA-TIRF) experiment at incidence angles of 68° and 73°, our results – slightly different from theirs – might be explained by the distinct cell types used, the smaller penetration depth in our case, or the fact that IRC images are difficult to interpret because the effects of RI and substrate-membrane spacing both contribute to the local image intensity.

Another way to measure the near-membrane RI is to use the RI-dependence of the critical angle for total internal reflection, but on the *excitation* site. In the past, several authors have noted the emergence of transmitted light propagating at very oblique angles at high-RI sites where TIR is disrupted [8, 16, 17]. Along the same lines, the steep dependence of the *reflected* intensity on the RI near the critical angle has been proposed to map the average RI from an angular scan of the reflected-beam intensity [18].

Other techniques that use the information contained in the reflected beam are digital holographic interferometry that has been implemented in a TIR geometry [19-21] (although no cellular measurements are reported here). Ash and co-workers later used this technique [22] to characterize *Amoeba proteus* as well as SKOV-3 cells (an ovary cancer cell line) and 3T3 fibroblasts.
Similarly, heterodyne interferometry uses the phase difference between *s*- and *p*-polarized components of totally reflected light [23] or the RI-dependent phase variation of *p*-polarized reflected light [24]. Related work uses the beam profile (intensity map) of the totally reflected beam of cells cultured on a gold-coated coverslip to measure the surface-plasmon resonance-generated contrast in reflection/absorption[25]. However, an absolute calibration of RI is missing.

On the emission side, Enderlein and co-workers used the BFP ring pattern to measure the effective NA of the microscope objective [26], an approach that we employed in our earlier work for characterizing the ZEISS objective used here [8] and that directly inspired the present study. In our present study, the procedure is just the inverse: instead of taking the BFP radii corresponding to the $\theta_{c,fluo}$ of water and air as a reference for calibrating $r_{max} = r_{NA,eff}$, we here used $r_{NA,eff}$ and a series of RI standards to calibrate BFP radii and took the radii measured for different organelle-specific labels for calculating the corresponding RI values. Building on the same idea, the Jaffiol lab proposed an approach for polar-incidence angle ($\theta$) calibration, based on the analysis of ring patterns on BFP images that indirectly uses the RI information encoded in the critical angle [27, 28]. Similarly, Soubies and co-workers use different RI standards to obtain a relation between the corresponding $\theta_c$s and BFP radii, which are then used for multi-angle TIRF microscope calibration [29]. These two studies are conceptually very close to ours in that they use the angle-radius correspondence, but they did not use the BFP information for cellular refractometry.

Using the dependency of the fluorescence lifetime on the local chemical microenvi-ronment [30] of genetically targeted GFP fusion proteins, van Manen *et al.* found RI values of ~1.38 and ~1.46 form cytosolic and membranous compartments, respectively [31]. Similar cytoplasmic values (1.36–1.39) were obtained with tomographic phase microscopy, with the RIs corresponding to subcellular organelles [32].

Generally, our RI values seem slightly smaller than published values (although a large scatter exists in the literature). This systematic difference could be due to the fact that in our case most dyes are exposed to mixed environments, i.e., lipid/protein-rich membrane on the organelle face but a more aqueous environment on the cytoplasmic face. Our values are generally closer to the mean cytoplasmic value (1.360 ± 0.004) found from quantitative phase-amplitude microscopy and confocal microscopy [33] than to values obtained from purified organelles. **Table S3** compiles published values from endo/lysosomal and mitochondrial RI measurements. We did not find RI values for the intact ER, but two older studies reported values ranging from 1.549 to 1.601 for 'endoplasmic vesicles' [34] and ~1.405-1.415 for the fractions #1-3 corresponding to the ER [35], respectively, after subcellular fractionation and ultracentrifugation.

# Supplementary references